\documentclass[prb]{revtex4}

 \usepackage{dcolumn,amsmath}
 \usepackage{bm}

\newcommand{\eref}[1]{(\ref{#1})}
\newcommand{\Eref}[1]{Eq.~(\ref{#1})}

\newcommand{\calA}{{\cal A}}
\newcommand{\calV}{{\cal V}}
\newcommand{\Titovref}{Titov:03Ae}
 \newcommand{\abinitio}{{\sl ab~initio} }

\begin{document}

\title{Accounting for correlations with core electrons
 by means of the generalized relativistic effective core potentials: 
 Atoms Hg and Pb and their compounds.}
\author{N.S.\ Mosyagin}\email{mosyagin@pnpi.spb.ru}
                       \homepage{http://www.qchem.pnpi.spb.ru}
\author{A.V.\ Titov}

\affiliation{Petersburg Nuclear Physics Institute, 
             Gatchina, St.-Petersburg district 188300, Russia}

\date{\today}

\begin{abstract}
 A way to account for correlations between the chemically active 
 (valence) and innermore (core) electrons in the framework of the generalized
 relativistic effective core potential (GRECP) method is suggested.  The
 ``correlated'' GRECP's (CGRECP's) are generated for the Hg and Pb atoms.  Only
 correlations for the external twelve and four electrons of them, 
 correspondingly, should be treated explicitly in the subsequent calculations 
 with these CGRECP's whereas the innermore electrons are excluded from the 
 calculations.  Results of atomic calculations with the correlated and 
 earlier GRECP versions are compared with the corresponding all-electron 
 Dirac-Coulomb values.  Calculations with the above GRECP's and CGRECP's are 
 also carried out for the lowest-lying states of the HgH molecule and its 
 cation and for the ground state of the PbO molecule as compared to earlier 
 calculations and experimental data. The accuracy for the vibrational 
 frequencies is increased up to an order of magnitude and the errors for 
 the bond lengths (rotational constants) are decreased in about two times 
 when the correlated GRECP's are applied instead of earlier GRECP versions 
 employing the same
 innercore-outercore-valence partitioning.
\end{abstract}


\maketitle

 \section{Introduction.}

 Accurate calculations of electronic structure of molecules are required in
 various fields of both basic research and practical applications.  To attain
 high accuracy, correlations not only for chemically active (valence and
 sometimes outermost core) but also for innermore electrons often have to be
 taken into account.  The number of the latters can be rather large (see atomic
 calculations below). For brevity, we will further refer in introduction 
 to these external chemically active and innermore core electrons as just to 
 the ``valence'' and ``core'' ones, correspondingly.

 It was studied, e.g., in papers\cite{Mosyagin:00,Isaev:00} for the Hg and Pb
 atoms that neglecting the correlations between the valence and core electrons
 leads to significant errors in calculating transition energies already between
 lowest-lying states.  The $5p$ and innermore shells of Hg as well as the $5d$
 and innermore shells of Pb were considered as the core shells in
 Refs.~\onlinecite{Mosyagin:00, Isaev:00} whereas the outermore shells were
 treated as valence. The correlations between the core and valence electrons
 which also include the terms of ``core-...-core-valence''-type in high orders
 of perturbation theory are called below as {\it core correlations}.  
 We will further consider only such a part of the core correlations which can
 be taken into account in a relativistic calculation of a free atom 
 with the codes fully exploiting spherical symmetry.
 The problem is that the computational efforts very fast grow with increasing
 the number of correlated electrons if the two-electron interactions are
 treated explicitly.  Therefore, the approximate methods which allow one to
 treat the core correlations by a simplified way are of considerable practical
 interest for accurate calculations. It is, obviously, reasonable to account 
 for the core correlations already at the stage of constructing effective
 atomic Hamiltonians if the separation of atomic shells on the valence and 
 core spaces is done appropriately (by analogy with the procedure of freezing 
 atomic core shells in molecular calculations, see section~\ref{sOC-Fr}).
 The question under consideration in the paper is concerning the accuracy of
 the large-core one-electron relativistic effective core potentials (RECP's) in
 simulating the energy-dependent effects of correlations between atomic core
 and valence electrons in heavy-atom molecules which include in general also
 two-electron and higher order core-valence interactions.

 The RECP method is widely used for calculations of molecules containing heavy
 atoms\cite{Ermler:88} because it reduces drastically the computational cost as
 compared to the all-electron four-component approach both at the integral
 generation--transformation stages and at the stage of correlation calculation
 when the spin-orbit basis set and two-step schemes of accounting for
 spin-orbit interaction etc.\ are used.  It is demonstrated both theoretically
 and computationally in Refs.~\onlinecite{Titov:99, Mosyagin:00, Isaev:00,
 Mosyagin:01b, Titov:01} that the RECP method can be used as a very accurate
 approximation not only for SCF calculations but for correlated calculations as
 well.  In a series of papers\cite{Titov:96b, Titov:99, Kozlov:97, Mosyagin:98,
 Petrov:02, Isaev:04, Petrov:04}, it is suggested to split a correlation
 calculation of a molecule containing heavy atoms onto computationally
 tractable consequent calculations in the valence and core regions, i.e.\ 
 molecular RECP calculation at the first step and one-center restoration of
 electronic structure in atomic cores at the second step.  In the two-step
 calculation, the computational efforts in correlating core and valence
 electrons will be roughly summed, whereas they have polynomial dependence on
 the number of explicitly treated electrons and on the basis set size in the
 one-step calculation since the number of varied parameters in correlation
 calculations (the numbers of coefficients in the configuration interaction or
 cluster amplitudes in the coupled cluster studies) grows proportionally to the
 number of the excitation operators used within a considered level of
 correlation treatment.  In the present paper, a method of treatment of the
 core correlations with the help of the generalized RECP (GRECP) operator is
 suggested.  We will further refer to GRECP's which account for such
 correlation effects as to the ``correlated'' GRECP's or CGRECP's. 

   Our first version of the CGRECP generation scheme was suggested in 
   Ref.~\onlinecite{Titov:99} and the corresponding CGRECP's were generated in 
   Ref.~\onlinecite{Titov:00}.  Unfortunately, this scheme had some drawbacks 
   and these CGRECP's were not published.  Later, our scheme for the CGRECP 
   generation was seriously improved and the results of the CGRECP calculations 
   for the Hg atom were first presented in table~5 in 
   Ref.~\onlinecite{Titov:03}.  In the present paper,
   calculations with our CGRECP's on the
    Hg and Pb atoms
    as well as
   on the HgH and PbO molecules are
    performed with a high level of accounting for correlation effects.

 It should be noted that the present CGRECP's and the core polarization
 potentials (CPP's) suggested in Refs.~\onlinecite{Mueller:84a, Mueller:84b,
 Fuentealba:82}
 simulate polarization-correlation effects by different ways.  The CPP's
 account directly for a displacement of core (dipole polarization) and
 radially asymmetric deformation of core (quadruple polarization etc.) 
 due to the electrostatic forces acting on the core from the valence 
 electrons, other cores and external electric fields.  However, they do not 
 account accurately for the 
 Pauli exclusion principle (requirement of orthogonality of electronic states 
 in core to occupied states of valence electrons and to core states of 
 neighbouring atoms) that leads to neglecting spin-polarization of core etc.
 Moreover, the CPP's do not take into account the effects of spherically
 symmetric relaxation of core due to the above described ``core correlations''.
 In turn, the present CGRECP's describe interactions between the spherically
 symmetric model of the correlated core and other electrons and nuclei.  These
 interactions, in particular, account for the core correlations and some
 ``spherically-averaged'' polarization effects.  (In more details, the
 one-electron part of the CPP operator accounts well for some part of the
 core-valence interactions (``vacuum polarization''-type terms) discussed above
 whereas its two-electron part accounts first of all for the
 valence-core-valence interactions (often called by ``screening''-type terms).
 The latter are not perfectly approximated within the present CGRECP's. 
 In turn, the one-electron CGRECP's account for the high-order correlation 
 effects within the core itself which are not described by CPP's.)

 Sufficiently accurate correlation method
 is required to describe the core correlations in \abinitio all-electron
 calculation, which are approximated then by the CGRECP (as one can conclude,
 e.g., from calculations of the Hg and Pb atoms\cite{Mosyagin:00, Mosyagin:01b,
 Isaev:00}).
 One can see from Tables~\ref{HgH} and \ref{PbO} of the present paper that 
 about one-half of the difference between the bond lengths calculated in the
 ``frozen core'' approximation and the experimental data
  can be described by the spherically-symmetric core correlation-relaxation.
  The incorporation of the GRECP and CPP methods to treat
  all the most important core correlation-relaxation (including polarization)
  effects simultaneously is suggested by us in future.

 In Ref.~\onlinecite{Maron:98}, only core polarization effects including mainly
 spin-orbit polarization and other ``electrostatic polarization effects on the
 core arising from single excitations out of the $(n-1)$ sub-valence shell''
 were taken into account by means of their multiconfigurational Dirac-Fock
 shape-consistent pseudopotentials (MCDF-SCPP's).  
 The spin-orbit polarization is
 automatically taken into account in our (and many other groups') RECP's
 because we generate two-component (spin-dependent) GRECP's on the base of
 all-electron reference calculations with the Dirac-Coulomb(-Breit)
 Hamiltonian.  Unlike Ref.~\onlinecite{Maron:98}, the present CGRECP's account
 for the core correlations.  In order to account for the latter by
 means of an \abinitio RECP, some reference \abinitio calculation should be
 performed in which the simulated correlations are explicitly considered.  When
 generating MCDF-SCPP\cite{Maron:98}, only one single $(n-1)p \rightarrow
 (n+1)p$ excitation from the main configuration is used in the reference
 all-electron calculations.  The polarization effects can be described in the
 all-electron calculation with the help of single excitations only 
 but at least double excitations are necessary to describe correlations
 satisfactorily.  Another question is concerning the possibility of the
 radially-local SCPP operator to reproduce the frozen core approximation with a
 good enough accuracy because the errors of the former can be, in principle,
 higher than the contribution of the core correlation-polarization effects.
 The forms of the operators should adequately describe the types of the
 described contributions whereas the errors of the radially-local SCPP
 approximation can not be properly compensated, e.g., by the CPP-type operator.
 The ways to increase accuracy of the RECP (SCPP) approximation are discussed
 in the next section.

\section{GRECP method.}
 \label{sGRECP}

 When core electrons of a heavy-atom molecule do not play an active role, the
 effective Hamiltonian with RECP can be presented in the form
\begin{equation}
   {\bf H}^{\rm Ef}\ =\ \sum_{i_v} \Bigl[{\bf h}^{\rm Schr}(i_v) +
          {\bf U}^{\rm Ef}(i_v)\Bigr] + \sum_{i_v > j_v} \frac{1}{r_{i_v j_v}}\ .
 \label{Heff2}
\end{equation}
 Hamiltonian \eref{Heff2} is written only for a valence subspace of electrons,
 which are treated explicitly and denoted by indices $i_v$ and $j_v$.  In
 practice, this subspace is often extended by inclusion of some outer core
 shells for better accuracy.  ${\bf U}^{\rm Ef}$ is an RECP (or relativistic
 pseudopotential) operator that can be written in the separable (e.g., see
 Ref.~\onlinecite{Theurich:01} and references) or radially-local
 (semi-local)\cite{Ermler:88} approximations when the valence pseudospinors are
 smoothed in heavy-atom cores.  Besides, the generalized RECP 
 operator\cite{Titov:99, Titov:00} 
 described below can be used that includes the
 radially-local, separable and Huzinaga-type\cite{Bonifacic:74} relativistic
 pseudopotentials as its components and some special cases.  Additionally, the
 GRECP operator can include terms of other types, called by ``self-consistent''
 and two-electron ``term-splitting'' corrections\cite{Titov:95, Titov:99,
 Titov:00}, which are important first of all for most economical (but precise)
 treatment of transition metals, lanthanides and actinides.  With these terms,
 accuracy provided by GRECP's can be even higher than the accuracy of the
 ``frozen core'' approximation (employing the same number of explicitly treated
 electrons) because they can account for relaxation of explicitly excluded
 (inner core) shells\cite{Titov:99}.  In~\Eref{Heff2}, ${\bf h}^{\rm Schr}$
 is the one-electron Schr\"odinger Hamiltonian

\begin{equation}
     {\bf h}^{\rm Schr}\ = - \frac{1}{2} {\vec \nabla}^2 
     - \frac{Z_{ic}}{r}\ ,
 \label{Schr}
\end{equation}
 where {$Z_{ic}$} is the charge of the nucleus decreased by the number of inner
 core electrons.  The (G)RECP operator simulates, in particular, interactions
 of the explicitly treated electrons with those which are excluded from the
 (G)RECP calculations.  Contrary to the four-component wave function used in
 Dirac-Coulomb(-Breit) calculations, the pseudo-wave function in the (G)RECP
 case can be both two- and one-component.  The use of the effective 
 Hamiltonian~\eref{Heff2} instead of the all-electron relativistic Hamiltonians 
 raises a
 question about its accuracy.  It was shown both theoretically and in many
 calculations (see Ref.~\onlinecite{Titov:99} and references) that a typical
 accuracy of the radially-local RECP versions is within 1000--3000~cm$^{-1}$
 for transition energies between low-lying states.

 The GRECP concept was introduced and developed in a series of papers (see
 Refs.~\onlinecite{Titov:99,Titov:00,Petrov:04b,Mosyagin:05a,Titov:05b} and 
 references).  In
 contrast to other RECP methods, GRECP employs the idea of separating the space
 around a heavy atom into three regions: inner core, outer core and valence,
 which are treated differently. It allows one to attain practically any desired
 accuracy, while requiring moderate computational efforts since the overall
 accuracy is limited in practice by possibilities of correlation methods.

\subsection{Generation of GRECP's with the separable correction.}
 \label{sGener} 

 The main steps of the scheme of generating the GRECP version with
 the separable correction taken into account are:
\begin{enumerate}
\item   The numerical all-electron relativistic calculation of a generator
        state is carried out for an atom under consideration. For this purpose,
        we use the {\sc hfdb} code\cite{Bratzev:77, Petrov:04b}
	for atomic calculations by Dirac-Fock~(DF) method
        (that can account also for the Breit effects self-consistently).
\item   The numerical pseudospinors  {$\tilde f_{nlj}(r)$} are
        constructed of the large components  {$f_{nlj}(r)$} of the
        outer core and valence DF bispinors so that the innermost
        pseudospinors of them (for each $l$ and $j$) are nodeless, the next
        pseudospinors have one node, and so forth. These pseudospinors satisfy
        the following conditions:
\begin{equation}
 \tilde f_{nlj}(r) =
 \left\{
  \begin{array}{ll}
   f_{nlj}(r),
                                 &  r\geq R_{c}, \\
   y(r)=
   r^{\gamma}\sum_{i=0}^{5}a_{i}r^{i}, &  r<R_{c},
  \end{array}
 \right.
 \label{tSmooth}
\end{equation}
\[
 \begin{array}{ccc}
  & l=0,1,\ldots,L,~~~~~ j=|l\pm\frac{1}{2}|, & \\
  & n=n_c,n_{c'},\ldots,n_v, &
 \end{array}
\]
        where $n_v, n_c, n_{c'}$ are principal quantum numbers of the
        valence and outer core spinors, {$L$} is one more than the highest
        orbital angular momentum of the inner core spinors. The leading
        power  {$\gamma$} in the polynomial is typically chosen to be close to
        {$L{+}1$} in order to ensure sufficient ejection of the valence and
        outer core electrons from the inner core region.  The $a_i$
        coefficients are determined by the following requirements: 
        \begin{itemize}
	\item    {$\{\tilde f_{nlj}\}$} set is orthonormalized\footnote{
           In practice, the orthogonality requirement of the outer core and
           valence pseudospinors often leads to rather singular potentials and
           should not be strictly satisfied at the stage of the potential
           generation. Some compromise between ``the most smooth''
           and ``the most accurate'' potentials should be chosen in
           order to use them efficiently in further molecular calculations.
           },
        \item    {$y$} and its first four derivatives match
		 {$f_{nlj}$} and its derivatives at $R_c$,
        \item    {$y$} is a smooth and nodeless function, and
        \item    {$\tilde f_{nlj}$} ensures a sufficiently
                smooth shape of the corresponding potential.
	\end{itemize}
         {$R_{c}$} is chosen near the extremum of the large component of the
         bispinor so that the corresponding pseudospinor has the defined above
         number of nodes. In practice, the  {$R_{c}$} radii for the different
         spinors should be chosen close to each other  to generate smooth
         potentials.
\item   The  {$U_{nlj}$} potentials are derived for each 
        {$l{=}0,\ldots,L$} and  {$j{=}|l \pm \frac{1}{2}|$}        
        for the valence and outer core pseudospinors so that
	the {$\tilde f_{nlj}$} are solutions of
	the nonrelativistic-type Hartree-Fock equations in the
	{\it jj}-coupling scheme for a ``pseudoatom'' with
        the removed inner core electrons.
\begin{eqnarray}
  U_{nlj}(r)  =  \tilde f_{nlj}^{-1}(r)
                \Biggl[\Biggl( \frac{1}{2} 
		{\bf \frac{d^{2}}{dr^{2}} }
                - \frac{l(l+1)}{2r^{2}}
                + \frac{Z_{ic}}{r} 
		 -   \widetilde{\bf J}(r) 
	          +  \widetilde{\bf K}(r)
       	  +  \varepsilon_{nlj}  \Biggr) \tilde f_{nlj}(r)
	   \nonumber\\
            +    \sum_{n'\neq n} \widetilde{\varepsilon}_{n'nlj}
                \tilde f_{n'lj}(r) \Biggr] 
          ,
 \label{U_nlj}
\end{eqnarray}
 where {$\widetilde{\bf J}$} and {$\widetilde{\bf K}$} are the
 Coulomb and exchange operators calculated with the
 {$\tilde f_{nlj}$} pseudospinors, {$\varepsilon_{nlj}$} are the
 one-electron energies of the corresponding bispinors, and
 {$\widetilde{\varepsilon}_{n'nlj}$} are off-diagonal Lagrange multipliers
 (which are, in general, slightly different for the cases of 
 the original bispinors and pseudospinors).
 
        In the case of the pseudospinor with nodes, the potential is singular
        because the zeros of the denominator and numerator, as a rule, do not
        coincide.  However, in practice, these zeros are close to each other as
        was demonstrated in Ref.~\onlinecite{Titov:91} and the most appropriate
        solution of this problem is interpolation of the potential in a
        vicinity of the pseudospinor node. The error of reproducing the
        one-electron energy due to such interpolation can be made small enough
        (because the pseudospinors are small in a vicinity of the node and the
        node position is not virtually changed at bond making and low-lying
        excitations). It does not exceed the errors of the GRECP approximation
        caused by smoothing the valence and outercore spinors
        and the approximate treating the interaction with the inner core
        electrons\cite{Tupitsyn:95}.
\item The GRECP operator with the separable correction written in the spinor
       representation\cite{Tupitsyn:95,Titov:99} is as
\begin{eqnarray}
 \label{UGRECP}
  {\bf U}^{\rm Ef}  &=&  U_{n_vLJ}(r)
                 +  \sum_{l=0}^L \sum_{j=|l-1/2|}^{l+1/2}
		   \Biggl\{\Bigl[U_{n_vlj}(r) 
		    -  U_{n_vLJ}(r)\Bigr]
                   {\bf P}_{lj}   \nonumber\\
                &+&   \sum_{n_c} 
                   \Bigl[U_{n_clj}(r) 
		   -  U_{n_vlj}(r)\Bigr] 
                   \widetilde{\bf P}_{n_clj} 
		   +   \sum_{n_c}  \widetilde{\bf P}_{n_clj}
                   \Bigl[U_{n_clj}(r) 
		    -  U_{n_vlj}(r)\Bigr] \\ 
              &-&   \sum_{n_c,n_{c'}} 
                   \widetilde{\bf P}_{n_clj}
                   \biggl[\frac{U_{n_clj}(r)+U_{n_{c'}lj}(r)}{2} 
		      -   U_{n_vlj}(r)\biggr]
                   \widetilde{\bf P}_{n_{c'}lj}\Biggr\}, \nonumber
\end{eqnarray}
 where
\[
  {\bf P}_{lj} = \sum_{m=-j}^j
    \Bigl| ljm \Bigl\rangle \Bigr\langle ljm \Bigr|,
\ \ \ \ \ \ \ \ \ \
  \widetilde{\bf P}_{n_clj} = \sum_{m=-j}^j
  \Bigl| \widetilde{n_cljm} \Bigl\rangle \Bigr\langle \widetilde{n_cljm} \Bigr|,
\]
         {$\bigl| ljm \bigl\rangle \bigr\langle ljm \bigr|$} 
 is the projector on the two-component spin-angular function 
	 {$\chi_{ljm}$}, 
         {$\bigl| \widetilde{n_cljm} \bigl\rangle
                \bigr\langle \widetilde{n_cljm} \bigr|$}
 is the projector on the outer core pseudospinors 
         {$\tilde f_{n_clj}\chi_{ljm}$}, 
         and  {$J=L+1/2$}.
\item   The numerical potentials and pseudospinors can be fitted
        by gaussian functions\cite{Mosyagin:97} to be used in calculations of
        polyatomic systems.
\end{enumerate}
 Two of the major features of the GRECP version with the separable correction
 described here are generating of the effective potential components for the
 pseudospinors which may have nodes, and addition of non-local separable terms
 with projectors on the outer core pseudospinors (the second and third lines 
 in~\Eref{UGRECP}) to the standard semi-local RECP operator (the first line 
 in~\Eref{UGRECP}). 
 These terms account for difference between potentials for outercore and 
 valence shells, which in $r>R_c$ is defined by smoothing within $R_c$ 
 as is shown in Ref.~\onlinecite{\Titovref} and in many cases this difference
 can not be neglected for ``chemical accuracy'' of valence energies.
 The more circumstantial description of distinctive features of
 the GRECP as compared to the original RECP schemes is given in
 Refs.~\onlinecite{Mosyagin:98c,Titov:00b}.  Some other GRECP versions are
 described and discussed in details in Refs.~\onlinecite{Titov:99, Titov:00,
 Mosyagin:05a}.

 The GRECP operator in the spinor representation \eref{UGRECP} is naturally
 used in atomic calculations.  The spin-orbit representation of this operator
 which can be found in Ref.~\onlinecite{Titov:99} is more efficient in practice
 being applied to calculation of molecules.  Despite the complexity of
 expression~\eref{UGRECP} for the GRECP operator, the calculation of its
 one-electron integrals is not notably more expensive than that for the case of
 the conventional radially-local RECP operator.

     \subsection{Freezing the innermost shells from the outer core space.}
\label{sOC-Fr}

 It was noted in Refs.~\onlinecite{Mosyagin:94,Tupitsyn:95} that using
 essentially different matching radii $R_c$ in \Eref{tSmooth} for different
 $lj$ is not expedient since the (G)RECP errors are mainly set by the
 outermost from them (see below).  In turn, explicit treatment of all of the
 outer core shells of an atom with the same principal quantum number is not
 usually reasonable in molecular (G)RECP calculations because of essential
 increase in computational efforts without serious improvement of accuracy.  A
 natural way out is to freeze the innermost of them before performing molecular
 calculation but this can not be done directly if the spin-orbit molecular
 basis set is used whereas the core shells should be better frozen as spinors.
 In order to exclude (``freeze'') explicitly those innermost shells (denoted by
 indices $f$ below) from molecular (G)RECP calculation without changing the
 radial node structure of other (outermore core and valence) shells in the core
 region, the energy level shift technique can be applied to overcome the above
 contradiction\cite{Titov:99, Titov:01}.  Following Huzinaga {\it et
 al.}\cite{Bonifacic:74}, one should add the effective core operator ${\bf
 U}^{\rm Ef}_{\rm Huz}$ containing the Hartree-Fock field operators, the
 Coulomb ($\tilde {\bf J}$) and spin-dependent exchange ($\tilde {\bf K}$)
 terms, over these core spinors together with the level shift terms to the
 one-electron part of the Hamiltonian:

\begin{equation}
  {\bf U}^{\rm Ef}_{\rm Huz} =
    \bigl({\bf \tilde J{-}\tilde K}\bigr)[\tilde f_{n_flj}]\ +
      \sum_{n_f,l,j} B_{n_flj}\
       \bigl|\tilde f_{n_flj} \bigr\rangle \bigl\langle \tilde f_{n_flj}\bigr|\
          \quad (\mbox{i.e.}\ \ \varepsilon_{n_flj} \to
                 \varepsilon_{n_flj}{+}B_{n_flj})\ ,
 \label{OC_Fr-1}
\end{equation}
 where the $B_{n_flj}$ parameters are of order $M|\varepsilon_{n_flj}|$ and
 $M > 1$ (usually $M \gg 1$ in our calculations).
 Such nonlocal terms are needed in order to prevent collapse of the
   valence electrons to the frozen core states. They introduce some ``soft
   orthogonality constraint'' between the ``frozen'' and outermore electronic
   states.

 All the terms with the frozen core spinors (the level shift operator
 and exchange interactions) can be transformed to the spin-orbit
 representation in addition to the spin-independent Coulomb term,
 using the identities for the ${\bf P}_{lj}$ projectors\cite{Hafner:79}:

\begin{equation}
        {\bf  P}_{l,j=l\pm 1/2}\
         =\ \frac{1}{2l{+}1} \Biggl[ \Bigl(l +
            \frac{1}{2} \pm \frac{1}{2}\Bigr)
            {\bf P}_l \pm
          2 {\bf P}_l\
          \vec{\bf l}{\cdot}\vec{\bf s}\ {\bf P}_l \Biggr]\ ,\ \
  {\bf P}_{l} =
     \sum\limits_{m_l=-l}^l \bigl| lm_l \bigl\rangle \bigr\langle lm_l \bigr|\ .
\label{Oper_Pnljs}
\end{equation}
 where $\vec{\bf l}$ and $\vec{\bf s}$ are operators of the orbital and spin
 momenta, {$\bigl| lm_l \bigl\rangle \bigr\langle lm_l \bigr|$} is the
 projector on the spherical function {$Y_{lm_l}$}.

 More importantly, these outer core pseudo{\it spinors} can be frozen in
 calculations with the {\it spin-orbit} basis sets and they can already be
 frozen at the stage of calculation of the one-electron matrix elements of the
 Hamiltonian, as implemented in the {\sc molgep} code\cite{MOLGEP}. Thus, any
 integrals with indices of the frozen spinors are completely excluded after the
 integral calculation step.  The multiplier $M=30$ was chosen in the present
 molecular calculations to prevent mixing the shifted core states to the
 wavefunction due to correlations but not to get poor reference wavefunction in
 the initial spin-averaged calculations at the same time (as would be for
 $M\to\infty$).

 The ``freezing'' of innermost shells from the outer core space within the
 ``small core'' GRECP's is sometimes required because the accuracy of the 
 GRECP's generated directly for a given number of explicitly treated electrons 
 cannot always correspond to the accuracy of the conventional frozen core
 approximation with the same space of explicitly treated electrons (without
 accounting for the frozen states). That space is usually chosen as a minimal
 one required for attaining a given accuracy. In fact, the ``combined'' GRECP,
 with separable and Huzinaga-type terms, is a new GRECP version treating the
 above number of electrons explicitly but which already provide the accuracy
 approaching to that of the frozen core approximation.  The efficiency of using
 the ``freezing'' procedure within the GRECP method was first demonstrated in
 calculations\cite{Titov:99,Titov:01} of Tl and TlH.

 Let us consider the advantages of using the combined GRECP version as compared
 to the conventional RECP's in reproducing the original core-valence
 interactions (correlation) on example of the Hg atom in more details.  It is
 clear, that at least the $5d$ shell of Hg should be explicitly treated in
 accurate calculations of molecules containing Hg (e.g., see
 Ref.~\onlinecite{Mosyagin:01b}).  For those calculations it would be optimal
 to use the RECP's with 12 electrons of Hg treated explicitly (12e-RECP's) such
 as the RECP of Ross {\it et al.}\cite{Ross:90} or our valence RECP
 version\cite{Tupitsyn:95}.  However, the explicit correlation of the outer
 core and valence electrons, occupying the $5d$ and $ns,np,nd$ ($n = 6,
 7,\dots$) orbitals, respectively, cannot be satisfactorily described in the
 framework of 12e-RECP's with nodeless $5d$, $6s$ and $6p$ pseudospinors,
 mainly because the smoothed valence pseudospinors have the wrong behaviour in
 the outer core region.  One-electron functions $\phi^{corr}_{x,k}(r)$, being
 some linear combinations of virtual orbitals, correlate to occupied orbitals
 $\phi^{occ}_x$ (where $x=c,v$ stands for the outer core and valence orbital
 indices) and are usually localized in the same space region as $\phi^{occ}_x$.
 Therefore, the original ``direct'' Coulomb two-electron integrals describing
 the outercore-valence correlation of $\phi^{occ}_c$ and $\phi^{occ}_v$ can be
 satisfactorily reproduced by those with the pseudoorbitals, despite their
 localization in different space regions.  However, a two-electron integral
 describing the ``exchange'' part of the outercore-valence correlation,
\begin{equation}
   \int_{\vec{r}}  d\vec{r}\
        \phi_{c,k'}^{{corr}^{\dagger}}(\vec{r})\phi^{occ}_v(\vec{r})\
   \int_{\vec{r}'} d\vec{r}'\
        \phi_{v,k}^{{corr}^{\dagger}}(\vec{r}')\phi^{occ}_{c}(\vec{r}')\
   \frac{1}{\vec{r} - \vec{r}'}\ ,
 \label{2elInt}
\end{equation}
 cannot be well reproduced because the valence pseudoorbitals are smoothed 
 in the outer core region where the outer core pseudoorbitals are localized 
 (for more theoretical details, see Ref.~\onlinecite{Titov:99}).

 To overcome this disadvantage, one should use RECP's with at least 20
 electrons, e.g., 20e-GRECP\cite{Mosyagin:97, Tupitsyn:95}.  In
 Ref.~\onlinecite{Tupitsyn:95}, for the case of the 20e-GRECP it was also shown
 that the $5s, 5p$ pseudospinors could be frozen while still providing
 significantly higher accuracy than 12e-RECP's because the valence and virtual
 $ns$ and $np$ ($n = 6,7,\dots$) pseudoorbitals in the former case already have
 the proper nodal structure in the outer core region.

 The freezing technique discussed above can be efficiently applied to those
 outer core shells for which the spin-orbit interaction is clearly more
 important than the correlation and relaxation effects. If the latter effects
 are neglected entirely or taken into account within ``correlated'' GRECP
 versions, the corresponding outer core pseudospinors can be frozen and the
 spin-orbit basis sets can be successfully used for other explicitly treated
 shells. This is true for the $5p_{1/2,3/2}$ subshells in Hg, contrary to the
 case of the $5d_{3/2,5/2}$ subshells. Freezing the outer core pseudospinors
 allows one to optimize an atomic basis set only for the orbitals which are
 varied or explicitly correlated in subsequent calculations, thus avoiding the
 basis set optimization for the frozen states and reducing the number of the
 calculated and stored two-electron integrals.  Otherwise, if the $5p$ shell
 should be correlated explicitly, a spinor basis set can be more appropriate
 than the spin-orbit one in a molecular calculation.

 As to the Pb atom, the use of nodeless pseudospinors for valence 
 $6s,6p,\ldots$ shells leads to large (G)RECP errors but ``freezing'' 
 $5s,5p,5d$ pseudospinors within 22-electron GRECP again gives 4-electron GRECP
 but with much smaller matching radii, therefore, its errors practically 
 concide with the errors of the ``frozen core'' approximation already.

 \section{Scheme of ``correlated'' GRECP generation.}

 The GRECP method was chosen to take into account the core correlations because
 it allows one to reproduce very accurately electronic structure in the valence
 region whereas the errors of the radially-local approximation of the RECP
 operator (or of the RECP's generated for only nodeless
  pseudospinors)\cite{Titov:99}
 as well as the errors of other approximations made in calculations can be more
 significant than the contributions from the core correlations (e.g., see
 Tables~VI and X of Ref.~\onlinecite{Fromager:04}).  To take account of the
 latter effects, we have chosen the Fock-space relativistic coupled cluster
 method with one- and two-body cluster amplitudes (FS~RCC-SD)\cite{Kaldor:04ba}
 because it has essential advantages in accounting for correlations with the
 core electrons whereas accurate treatment of correlations between valence
 electrons is not so important for the CGRECP generation stage.  This method is
 size-consistent that is, in particular, significant for the compounds of heavy
 elements.
 The FS~RCC computational scheme, in which the part of correlations from lower
 sectors is ``frozen'' in the higher Fock space sectors,
 is especially suitable for incorporating the most important correlations of
 them into CGRECP.  Neglecting the higher order cluster amplitudes seems us
 reasonable because the core correlations give relatively small corrections to
 the properties determined mainly by the valence electrons.  This approximation
 can be compared to that made in the scheme of constructing conventional
 GRECP's on the base of the Dirac-Fock(-Breit) calculations despite these
 GRECP's are suggested to be used in accurate correlation calculations, see
 Ref.~\onlinecite{Titov:99} for theoretical details. At last, the atomic {\sc
 rcc-sd} code\cite{Kaldor:04ba} is very efficient because it fully exploits the
 spherical symmetry of atoms.

\vspace{3mm}
 The main steps of the current scheme of generation of the CGRECP's are as
 follows:
\begin{enumerate}
\item
     For a considered atom, a set of occupied spinors is derived from an
     all-electron DF calculation for some closed shell state which is
     energetically close to the states of primary interest in calculations with
     the constructed CGRECP. The unoccupied spinors are obtained with the help
     of some procedure for a basis set generation (e.g., described in
     Refs.~\onlinecite{Mosyagin:00, Isaev:00}).  Other basis set generation
     procedures could be also applied at this step because very large basis
     sets can be used in atomic calculations unlike the following molecular
     (G)RECP calculations.  The Fock matrix and two-electron integrals are
     calculated in this basis set with all-electron Dirac-Coulomb or
     Dirac-Coulomb-Breit Hamiltonians.
\item
     Two equivalent FS~RCC-SD calculations are carried out with the same 
     spaces of active spinors\cite{Paldus:03} and schemes of calculation.
In the present work, the closed shell ground states of 
the Hg$^{2+}$ and Pb$^{2+}$ ions served as references and the Fock-space 
schemes were
\begin{equation}
 {\rm Hg}^{3+} \leftarrow {\rm Hg}^{2+} \rightarrow {\rm Hg}^+ , ~~~
 {\rm Pb}^{3+} \leftarrow {\rm Pb}^{2+} \rightarrow {\rm Pb}^+ ,
\label{FS1}
\end{equation}
 with electrons added in the $6s_{1/2},6p_{1/2},6p_{3/2},6d_{3/2},6d_{5/2},
 5f_{5/2},5f_{7/2},5g_{7/2},5g_{9/2}$ and $6p_{1/2},6p_{3/2}$ active spinors 
 of Hg and Pb, respectively, and removed from the $5d_{3/2},5d_{5/2}$ Hg and 
 $6s_{1/2}$ Pb spinors.
     Only valence electrons are correlated in calculation~$\calV$, whereas both
     the valence and core electrons are correlated in calculation~$\calA$.  The
     active space should contain the spinors, for which the CGRECP components
     will be then constructed at step~{\it(5)}.
     The $6d,5f,5g$ CGRECP components for Pb were constructed employing the
     conventional GRECP generation scheme with the $6s$ and innermore shells
     ``frozen'' after step~{\it(5)}.  It was checked on example of Hg that this
     simplification of the generation procedure leads to negligible changes in
     the results of the CGRECP calculations.  As a result of the FS~RCC-SD
     calculations, a set of the one-body ($t_i^a$) and two-body ($t_{ij}^{ab}$)
     cluster amplitudes and ionization potentials ($e_m$) or electron
     affinities ($e_v$) is obtained. The $m$ and $v$ indices run over the
     active spinors occupied and unoccupied in the starting closed shell state,
     correspondingly.  The $i,j$ indices in the cluster amplitudes run over the
     spinors occupied in the above closed shell state and can, additionally,
     include the $v$ indices.  The $a,b$ indices in the cluster amplitudes run
     over the spinors unoccupied in the above closed shell state and can,
     additionally, include the $m$ indices.  If the correlations of the
     electron in state $i$ are not considered (e.g., in calculation~$\calV$),
     we put the corresponding $t_i^a$, $t_{ji}^{ab}$ and $t_{ij}^{ab}$ cluster
     amplitudes to zero. 
\item
     Differences $\Delta t_i^a=t_i^a[\calA]-t_i^a[\calV]$, $\Delta
     t_{ij}^{ab}=t_{ij}^{ab}[\calA]-t_{ij}^{ab}[\calV]$, $\Delta
     e_m=e_m[\calA]-e_m[\calV]$ and $\Delta e_v=e_v[\calA]-e_v[\calV]$ are 
     calculated.
     If the absolute values of the $\Delta t_i^a$, $\Delta e_m$ and $\Delta
     e_v$ differences are less than some threshold ($10^{-6}$ in the present
     work), we go to step~{\it(5)}. If they are not, we go to step~{\it(4)}.
     In the present generation scheme, the $\Delta t_{ij}^{ab}$ differences are
     neglected. However, they could be later compensated with the help of the
     two-electron term-splitting correction\cite{Titov:99} for higher accuracy.
\item
     We use $\Delta t_i^a$ to rotate the spinors ($\phi$) in the basis set 
\begin{equation}
   \phi_i = \phi_i^{\rm prev} + \sum_a \Delta t_i^a \phi_a^{\rm prev}~~~~~ 
   \mbox{ for~~~} i\not\in\{v\} \mbox{~~~and~~~} a\not\in\{m\},
\end{equation}
\begin{equation}
   \phi_v = \phi_v^{\rm prev} + \sum_a \Delta t_v^a \phi_a^{\rm prev},~~~~~
   \phi_m = \phi_m^{\rm prev} - \sum_i \Delta t_i^m \phi_i^{\rm prev},
\end{equation}
 where $\phi^{\rm prev}$ is the spinors obtained at the previous iteration.
 The derived spinors are then orthonormalized by the Schmidt procedure. The
 Fock matrix and two-electron integrals are calculated in the obtained basis
 set.  We use $\Delta e_m$ and $\Delta e_v$ to modify the diagonal Fock matrix
 elements ($F_{mm}$ and $F_{vv}$) only for calculation~$\calV$
\begin{equation}
   F_{mm}[\calV]=F_{mm}^{\rm prev}[\calV]-\Delta e_m,~~~~~
   F_{vv}[\calV]=F_{vv}^{\rm prev}[\calV]-\Delta e_v,
\end{equation}
     where $F^{\rm prev}[\calV]$ is the Fock matrix derived at the previous
     iteration.  We put the nondiagonal Fock matrix elements for 
     calculation~$\calV$ to zero. Then, we go to step~{\it(2)}.
\item
     The spinors and the corresponding Fock matrix elements from 
     calculation~$\calV$ are used instead of the original spinors and their 
     one-electron energies at the CGRECP generation scheme employing 
     the procedure of the GRECP generation described in section~\ref{sGener}.
\end{enumerate}

 Some of the most important properties of the generation schemes of reliable
 RECP's are their ``basis-set-independence'' and
 ``correlation-method-independence''. It means that extension of a one-electron
 basis and improving the level of the correlation treatment should not lead to
 decreasing accuracy of calculations with the used RECP.  These properties do
 not always take place for the well-known RECP's (pseudopotentials) but they
 are, as a rule, fulfilled for the shape-consistent RECP's and GRECP's.  The
 case of the correlated RECP's including CGRECP's is, obviously, more critical.
 Nevertheless, the discussed scheme of the CGRECP generation seems us flexible
 enough in these aspects since large basis sets can be employed within the
 FS~RCC-SD calculations of atoms (in particular, 
 their quality is sufficient for the accuracy of our interest in generating the
 CGRECP's for Hg and Pb, see below) whereas the contribution from three- and
 higher-body RCC amplitudes to the core correlation effects is not expected to
 be essential.

 \section{Atomic calculations.}

 Correlation structure of the Hg and Pb atoms was studied accurately in
 Refs.~\onlinecite{Mosyagin:00,Isaev:00}. It can be seen from
 Ref.~\onlinecite{Mosyagin:00} that at least 34 external electrons of Hg should
 be correlated 
 if consistent agreement with experimental data better than 200~cm$^{-1}$ for
 energies of one-electron transitions is desired for low-lying states.  Such
 accuracy is of practical interest in many cases (for chemistry of
 $d,f$-elements, many oxides etc.).  Then, it was shown in
 Ref.~\onlinecite{Mosyagin:01b} that the three-body cluster amplitudes for 12
 external electrons of the mercury atom and 13 electrons of the mercury hydride
 molecule are required to obtain accurate results.  Moreover, the 
 polarization/relaxation of the $5d$ shell of Hg is rather large
 in chemical bonding and should be treated explicitly.
 We describe correlations with the ``spherically symmetric'' $4f,5s,5p$ core
 shells of Hg by means of the CGRECP whereas the correlations for the
 $5d,6s,6p$ shells should be taken into account explicitly in the following
 calculations with this CGRECP.  For Pb we include the correlations with the
 $4s,4p,4d,4f,5s,5p,5d$ shells into the CGRECP.  It is in agreement with
 chemical intuition since Hg is a transition metal whereas the $5d$ shell in Pb
 is not so active chemically.  It should be noted that the core-valence
 partitioning used for incorporating the correlation effects into CGRECP
 differs from the ``innercore-outercore-valence'' partitioning used in the
 conventional GRECP generation procedure.  Therefore, those core shells which
 are explicitly treated with a GRECP constructed within the conventional scheme
 but whose correlations are taken into account at the CGRECP generation stage
 must be considered as ``frozen'' in the subsequent CGRECP calculations (see
 section~\ref{sOC-Fr} and the following section for details).  The states used
 in the FS~RCC-SD calculations at step~{\it(2)} of the CGRECP generation
 (``generator states'') are
  $5d_{3/2}^4 5d_{5/2}^6$, $5d_{3/2}^3 5d_{5/2}^6$, $5d_{3/2}^4 5d_{5/2}^5$,
  $5d_{3/2}^4 5d_{5/2}^6 6s_{1/2}^1$, $5d_{3/2}^4 5d_{5/2}^6 6p_{1/2}^1$,
  $5d_{3/2}^4 5d_{5/2}^6 6p_{3/2}^1$,
  $5d_{3/2}^4 5d_{5/2}^6 6d_{3/2}^1$,  $5d_{3/2}^4 5d_{5/2}^6 6d_{5/2}^1$, 
  $5d_{3/2}^4 5d_{5/2}^6 5f_{5/2}^1$,  $5d_{3/2}^4 5d_{5/2}^6 5f_{7/2}^1$, 
  $5d_{3/2}^4 5d_{5/2}^6 5g_{7/2}^1$,  $5d_{3/2}^4 5d_{5/2}^6 5g_{9/2}^1$ 
  for Hg and
  $6s_{1/2}^2$, $6s_{1/2}^1$, $6s_{1/2}^2 6p_{1/2}^1$, $6s_{1/2}^2 6p_{3/2}^1$
 for Pb. The gaussian expansions of the CGRECP's generated following the above
 discussed scheme as well as of the earlier GRECP versions for Hg and Pb can
 be found on our website http://www.qchem.pnpi.spb.ru/GRECPs\,.

 To check accuracy of the generated CGRECP's, we carried out comparative RCC
 and configuration interaction (CI) calculations, 
 in which the correlations were explicitly considered only
 for the valence electrons.  Both CGRECP and earlier GRECP versions were used
 in these calculations, in which the core shells were treated as frozen.  We
 will further refer to these earlier GRECP versions with the frozen core shells
 as to VGRECP's (in molecular calculations these shells are ``frozen'' within
 the VGRECP's {\it directly}, see section~\ref{sOC-Fr} and the next section for
 details).  The obtained results are presented in Tables~\ref{Hg} and \ref{Pb}
 and they are also compared with the results of the corresponding calculations
 (carried out by us earlier\cite{Mosyagin:00,Isaev:00}) when employing the
 all-electron Dirac-Coulomb Hamiltonian, in which the correlations were taken
 into account for both the valence and core electrons. The contributions from
 the core correlations to energies of transitions with excitation or ionization
 of a single electron are up to 1100~cm$^{-1}$ for Hg and 1000~cm$^{-1}$ for
 Pb.  One can also see that the CGRECP's allow one to reproduce the results of
 the corresponding Dirac-Coulomb calculations with accuracy better than
 310~cm$^{-1}$ for Hg and 390~cm$^{-1}$ for Pb that is on the level of so
 called ``chemical accuracy'' (about 1~kcal/mol or 350~cm$^{-1}$). Similar
 precision can be expected in calculations of at least vertical excitation
 energies for their compounds.

 \section{Molecular calculations.}

 The correlation structure for the lowest-lying states of the HgH molecule and
 HgH$^+$ ion was carefully studied in Ref.~\onlinecite{Mosyagin:01b} with the
 help of the {\sc rcc-sd} and nonrelativistic (one-component) {\sc cc-sdt}
 codes\cite{RCCSD, CCSDT}  (the latter was used in the scalar-relativistic RCC
 calculations to account for three-body cluster amplitudes).  A good agreement
 with the experimental data was attained when 19 external electrons of the HgH
 molecule were correlated by the RCC-SD method\cite{Kaldor:04ba} (19e-RCC-SD)
 with an approximate accounting for the contribution of the three-body cluster
 amplitudes for 13 outermost electrons and the counterpoise
 correction\cite{Gutowski:86,Liu:89} applied.  The above numbers of electrons
 are, obviously, smaller by one for the HgH$^+$ ion.  The conventional GRECP
 version from Ref.~\onlinecite{Mosyagin:97} was used in
 calculations\cite{Mosyagin:01b}.  The aim of the calculations presented in
 Table~\ref{HgH} is to study how the results of the GRECP/19e-RCC-SD
 calculation will be reproduced in the 13e-RCC-SD calculation with the CGRECP
 generated for Hg in the present work.  Twelve electrons are explicitly
 considered in calculations with both the VGRECP and CGRECP because the $5s,5p$
 pseudospinors of Hg are described within them by adding the
 Huzinaga-potential-type terms to the conventional GRECP operator (in some
 sense the core {\it spinors} are ``frozen'' with the help of the level-shift
 technique even when the {\it spin-orbit} basis set is employed for the valence
 electrons, see section~\ref{sOC-Fr}).
 The same basis set as in Ref.~\onlinecite{Mosyagin:01b} was used for the
 present VGRECP and CGRECP calculations.  The other details of these
 calculations are described in Ref.~\onlinecite{Mosyagin:01b} where they were
 designated as the RCC-SD-1 ones.

 One can see from Table~\ref{HgH} that application of the CGRECP instead of the
 VGRECP in the 13e-RCC calculation allows one to improve the agreement with the
 GRECP/19e-RCC calculation for the vibrational frequencies up to an order of
 magnitude.  The improvement up to two times is observed for the bond lengths
 (rotational constants). The errors for the dissociation energies are decreased
 but the errors for the transition energies are increased.  However, both the
 errors are on the level of ``chemical accuracy''. The higher-order Dunham
 coefficients ($w_e x_e$, $\alpha_e$, $-Y_{02}$) are reproduced, as a rule,
 worse with the CGRECP than with the VGRECP.  It should be noted that the
 correlations with the $4f$ and $5s$ shells of Hg were neglected in the
 GRECP/19e-RCC calculations. In turn, the CGRECP for Hg takes into account
 these correlations.  Therefore, the best suitable calculation to estimate the
 accuracy of the CGRECP is the 35e-RCC one.  It is shown in
 Ref.~\onlinecite{Mosyagin:00}, however, that the contributions from the
 correlations with the $4f$ and $5s$ shells mainly cancel each other.

 The PbO molecule is of interest now first of all in connection with the
 ongoing and suggested experiments on search for the electric dipole moment of
 the electron on the excited $a(1)$\cite{DeMille:00} and 
 $B(1)$\cite{Egorov:01} states.  In particular, calculations\cite{Isaev:04,
 Petrov:04} of the effective electric field, $W_d$, seen by an unpaired
 electron are necessary for interpretation of the experimental results.
 Calculation of spectroscopic properties can be useful to search for some
 better scheme of populating the working state.  The aim of the present
 calculations is to check accuracy and reliability of the CGRECP version used
 by us on example of the ground state of PbO, for which highly accurate
 experimental data are available.

 We carried out 10-electron spin-orbit direct CI (10e-SODCI)\cite{Buenker:99,
 Alekseyev:04a} calculations of the potential curve for the ground state of the
 PbO molecule (see Refs.~\onlinecite{Titov:01, Mosyagin:02, Isaev:02} for the
 details of such calculations). The calculations were carried out for 20
 internuclear distances from 2.7~a.u.\ to 4.6~a.u.\ with interval of 0.1~a.u.
 Molecular spectroscopic constants were calculated by the Dunham method in the
 Born-Oppenheimer approximation using the {\sc dunham-spectr}
 code\cite{Mitin:98}.  The core of Pb was simulated with the help of the VGRECP
 version from Refs.~\onlinecite{Mosyagin:97, Titov:99} and with the CGRECP
 version generated in the present work.  Four electrons are explicitly
 considered in calculations with both the VGRECP and CGRECP, in which the
 $5s,5p,5d$ pseudospinors of Pb are ``frozen'' with the help of the level-shift
 technique (see section~\ref{sOC-Fr}).  No relativistic effects were accounted
 for oxygen and, besides, its $1s$ shell was frozen in the PbO calculations.
 Thus, only four electrons of lead and six electrons of oxygen were explicitly
 correlated.  The same basis set was used for the VGRECP and CGRECP
 calculations.  We estimated that the counterpoise corrections for the $^3P_0$
 and $^3P_2$ states of Pb and for the $^3P$ state of oxygen are less than
 0.02~eV for all the considered internuclear distances.  The calculations at
 higher level of accuracy were not carried out because the corresponding SODCI
 calculations became very consuming.  Therefore, the counterpoise corrections
 were neglected when the spectroscopic constants were calculated.  The obtained
 results are presented in Table~\ref{PbO}. One can see that both results are in
 reasonably good agreement with the experimental data.  The largest improvement
 (about six times) for the CGRECP with respect to VGRECP is observed for the
 vibrational frequencies. Similar improvement was also observed for the
 frequencies in HgH.  The errors for the dissociation energy (approximately
 2~kcal/mol for breaking the double bond) are comparable by magnitude for the
 VGRECP and CGRECP. These errors can be partly explained by absence of the
 $h$-type functions in the used basis sets, insufficient level of accounting
 for the correlation in the above restricted CI calculations, etc.  The
 agreement for other spectroscopic constants with the experimental data is
 improved in about two times when the CGRECP is applied instead of the VGRECP.
 Similar result was also obtained for the bond lengths in HgH.

\section*{ACKNOWLEDGMENTS.}

 We are grateful to R.J.Buenker and his colleagues H.-P.Liebermann and
 A.B.Alekseyev for giving us the new version of {\sc sodci} code and to
 U.Kaldor and E.Eliav for the {\sc rcc-sd} code which were used in the present
 calculations.  The present work is supported by U.S.\ CRDF Grant No.\
 RP2--2339--GA--02 and by the RFBR grant 03--03--32335.  N.M.\ is also
 supported by the grants of Russian Science Support Foundation and the governor
 of Leningrad district.
 A part of the CI calculations was performed on computers of Boston University
 in the framework of the MARINER project.

\bibliographystyle{apsrev}

\bibliography{bib/JournAbbr,bib/Titov,bib/TitovLib,bib/Kaldor,bib/Isaev,bib/AbsConf/TitovAbs}

\begin{table}
\caption{The experimental transition energies between the low-lying states of 
   the mercury atom and its cations are taken from Ref.~\onlinecite{Moore:58}.
   The absolute errors in the transition energies from 
   12-electron\protect\footnote{This number is smaller by one or two for Hg$^+$
   or Hg$^{2+}$ ions, respectively.} FS~RCC-SD calculations with the VGRECP, 
   CGRECP and all-electron Dirac-Coulomb (DC) Hamiltonian in the 
   $[7,9,8,6,7,7]$ correlation basis set from Ref.~\onlinecite{Mosyagin:00} 
   are calculated with respect to the transition energies from 34-electron 
   FS~RCC-SD calculation with all-electron DC Hamiltonian. All values are 
   in cm$^{-1}$.}
\begin{tabular}{lrrrrr}
\hline
\hline
 State (leading & \multicolumn{2}{c}{Transition energies} & \multicolumn{3}{c}{Absolute errors with respect to DC/34e-RCC}\\
 conf., term)   & Exper.\ & DC/34e-RCC\cite{Mosyagin:00} & DC/12e-RCC\cite{Mosyagin:00} & VGRECP/12e-RCC & CGRECP/12e-RCC \\
\hline                                                                  
 $5d^{10} 6s^2 (^1S_0) \rightarrow$ & &        &       &       &     \\
 $5d^{10} 6s^1 6p^1 (^3P_0)$ &  37645 &  37471 &  -263 &  -229 & 305 \\
 $5d^{10} 6s^1 6p^1 (^3P_1)$ &  39412 &  39318 &  -326 &  -293 & 296 \\
 $5d^{10} 6s^1 6p^1 (^3P_2)$ &  44043 &  44209 &  -534 &  -497 & 264 \\
 $5d^{10} 6s^1 6p^1 (^1P_1)$ &  54069 &  55419 &  -650 &  -628 &  96 \\
 $5d^{10} 6s^1 (^2S_{1/2})$  &  84184 &  84550 &  -665 &  -640 & 206 \\
 $5d^{10} 6s^1 (^2S_{1/2}) \rightarrow\!\!$ & &    &       &       &     \\
 $5d^{10} 6p^1 (^2P_{1/2})$  &  51485 &  52025 &  -510 &  -465 &  35 \\
 $5d^{10} 6p^1 (^2P_{3/2})$  &  60608 &  61269 &  -793 &  -738 &  52 \\
 $5d^{10} (^1S_0)$           & 151280 & 151219 & -1087 & -1045 &  44 \\
\hline
\hline
\end{tabular}
\label{Hg}
\end{table}

\begin{table}
\caption{The experimental transition energies between low-lying electronic 
   states of the lead atom are taken from Ref.~\onlinecite{Moore:58}.  The
   absolute errors of all-electron Dirac-Coulomb (DC), VGRECP and CGRECP
   calculations with the help of the RCC and CI methods for 4 and 36 explicitly
   correlated electrons in the $[7,6,6,4,4]/[3,5,3,2]$ basis set from
   Ref.~\onlinecite{Isaev:00} are calculated with respect to the experimental
   data. All values are in cm$^{-1}$.}
\begin{tabular}{lrrrrr}
\hline
\hline
 State (leading                   &         & \multicolumn{4}{c}{Absolute errors with respect to the experimental data}\\
 conf., term)                     & Exper.\ & DC/4e-CI\cite{Isaev:00}   
		                                       & VGRECP/4e-CI   
                                                                 & DC/36e-RCC{+}VCIC\protect\footnote{36 electron FS~RCC-SD
calculation with valence CI Correction (VCIC)\cite{Isaev:00} as the difference in the total energies
from 4e-CI and 4e-FS~RCC-SD calculations accounting approximately for the three- and four-body 
cluster amplitudes for four valence electrons.}    
				                                            & CGRECP/4e-CI   \\
\hline
 $6s_{1/2}^2 6p_{1/2}^2 (^3P_0) \rightarrow$ &         &          &         &          &         \\ 
 $6s_{1/2}^2 6p_{1/2}^1 6p_{3/2}^1 (^3P_1) $ &    7819 &     -807 &    -740 &      -60 &    -114 \\ 
 $6s_{1/2}^2 6p_{1/2}^1 6p_{3/2}^1 (^3P_2) $ &   10650 &     -752 &    -668 &      157 &      89 \\ 
\hline                                                                                    
 $6s_{1/2}^2 6p_{3/2}^2 (^1D_2)            $ &   21457 &    -1707 &   -1619 &       26 &    -128 \\ 
 $6s_{1/2}^2 6p_{3/2}^2 (^1S_0)            $ &   29466 &    -1553 &   -1430 &      100 &     335 \\ 
\hline
\hline
\end{tabular}
\label{Pb}
\end{table}

\begin{table}
\caption{ Spectroscopic constants of the lowest-lying states of 
   the HgH molecule and HgH$^+$ ion from 13 and 19
   electron\protect\footnote{These numbers are smaller by one for the HgH$^+$
   ion.}  RCC-SD calculations with the GRECP, VGRECP and CGRECP in the H
   $(8,4,3)/[4,2,1]$ ANO and Hg $(14,12,9,3,2)/[7,7,4,2,1]$ basis set
   from Ref.~\onlinecite{Mosyagin:01b}.  
   All the results are corrected by counterpoise
   corrections calculated for the Hg $6s^2$ state.
   $R_e$ is in \AA, $D_e$ in eV, $Y_{02}$ in $10^{-6}$~cm$^{-1}$, 
   other values in cm$^{-1}$.
   }
\begin{tabular}{lccccccc}
\hline
\hline
HgH ($\sigma^2\sigma^1$) $^2\Sigma_{1/2}^+$ 
                & $R_e$ & $w_e$ & $D_e$ & $B_e$ & $w_e x_e$ & $\alpha_e$ & $-Y_{02}$ \\
\hline
 VGRECP/13e-RCC & 1.709 & 1575 &  0.35 & 5.76 &      56 &    0.262 &     312 \\
 CGRECP/13e-RCC & 1.705 & 1595 &  0.34 & 5.78 &      59 &    0.265 &     308 \\
  GRECP/19e-RCC & 1.702 & 1597 &  0.34 & 5.80 &      56 &    0.259 &     310 \\
\hline
HgH$^+$ ($\sigma^2$) $^1\Sigma_0^+$ 
                & $R_e$ & $w_e$ & $D_e$ & $B_e$ & $w_e x_e$ & $\alpha_e$ & $-Y_{02}$ \\
\hline
 VGRECP/12e-RCC & 1.596 & 2037 &  2.67 & 6.60 &      39 &    0.200 &     279 \\
 CGRECP/12e-RCC & 1.593 & 2070 &  2.73 & 6.63 &      40 &    0.200 &     273 \\
  GRECP/18e-RCC & 1.588 & 2067 &  2.72 & 6.66 &      39 &    0.199 &     278 \\
\hline
HgH$^*$ ($\sigma^2\pi^1$) $^2\Pi_{1/2}$ 
                & $R_e$ & $w_e$ & $T_e$ & $B_e$ & $w_e x_e$ & $\alpha_e$ & $-Y_{02}$ \\
\hline
 VGRECP/13e-RCC & 1.586 & 2067 & 24157 & 6.68 &      39 &    0.203 &     281 \\
 CGRECP/13e-RCC & 1.583 & 2104 & 24324 & 6.71 &      41 &    0.203 &     274 \\
  GRECP/19e-RCC & 1.578 & 2100 & 24044 & 6.75 &      39 &    0.201 &     280 \\
\hline
HgH$^*$ ($\sigma^2\pi^1$) $^2\Pi_{3/2}$ 
                & $R_e$ & $w_e$ & $T_e$ & $B_e$ & $w_e x_e$ & $\alpha_e$ & $-Y_{02}$ \\
\hline
 VGRECP/13e-RCC & 1.583 & 2086 & 27691 & 6.71 &      38 &    0.198 &     279 \\
 CGRECP/13e-RCC & 1.580 & 2123 & 27951 & 6.73 &      39 &    0.197 &     272 \\
  GRECP/19e-RCC & 1.576 & 2117 & 27629 & 6.77 &      37 &    0.197 &     278 \\
\hline
\hline
\end{tabular}
\label{HgH}
\end{table}

\begin{table}
\caption{Spectroscopic constants for the ground state of the PbO molecule 
   from 10-electron SODCI calculations with the VGRECP and CGRECP in the Pb
   $(22,18,12,6,5)/[5,6,4,3,1]$ and O $(14,9,4,3,2)/[6,5,4,3,2]$ basis set.
   $R_e$ is in \AA, $D_e$ in eV, $\alpha_e$ in $10^{-3}$~cm$^{-1}$, $Y_{02}$ 
   in $10^{-6}$~cm$^{-1}$, other values in cm$^{-1}$.
}
\begin{tabular}{lccccccc}
\hline
\hline
 & $R_e$ & $w_e$       & $D_e$ & $B_e$       & $w_e x_e$   & $\alpha_e$            & $-Y_{02}$             \\
\hline
 VGRECP/10e-SODCI           & 1.943 & 715 & 3.79 & 0.301 & 3.27 & 1.73 & 0.213 \\
 CGRECP/10e-SODCI           & 1.933 & 720 & 3.77 & 0.304 & 3.42 & 1.84 & 0.216 \\
 Experiment\cite{Huber:79} & 1.922 & 721 & 3.87 & 0.307 & 3.54 & 1.91 & (0.223)\protect\footnote{Cited in Ref.~\onlinecite{Huber:79} as uncertain.}\\
\hline
\hline
\end{tabular}
\label{PbO}
\end{table}

\end{document}